\documentclass[useAMS, usenatbib, usegraphicx]{mn2e}
\usepackage{pslatex,color}
\usepackage{bm,amsmath,amsfonts,amssymb,amsxtra,url}
\usepackage[utf8]{inputenc}
\usepackage[T1]{fontenc}
\usepackage{graphicx, natbib}

\usepackage{color}

\DeclareMathOperator{\arctanh}{arctanh}
\DeclareMathOperator{\sech}{sech}

\def\be{\begin{equation}}
\def\ee{\end{equation}}
\def\bea{\begin{eqnarray}}
\def\eea{\end{eqnarray}}

\newcommand{\PR}[1]{\ensuremath{\left[#1\right]}}
\newcommand{\PC}[1]{\ensuremath{\left(#1\right)}}
\newcommand{\chaves}[1]{\ensuremath{\left\{#1\right\}}}

\graphicspath{{./figures/}}

\title[Phenomenological and theoretical DM profiles] 
{Testing phenomenological and theoretical models of dark matter density profiles with galaxy clusters}
\author[Leandro J. {Beraldo e Silva}, Marcos Lima and Laerte Sodré Jr.]
{Leandro J. {Beraldo e Silva}$^{1,2,3}$\thanks{E-mail: lberaldo@if.usp.br}, 
Marcos Lima$^{1}$\thanks{E-mail: mlima@fma.if.usp.br} 
and Laerte {Sodré} Jr.$^{4}$\thanks{E-mail: laerte@astro.iag.usp.br}\\
$^{1}$ Departamento de F\'isica Matem\'atica, Instituto de Física, Universidade de São Paulo, S\~ao Paulo SP, Brazil\\
$^{2}$ Institut d'Astrophysique de Paris, Paris, France\\
$^{3}$ CAPES Foundation, Ministry of Education of Brazil, Bras\'ilia - DF 70.040-020, Brazil\\
$^{4}$ Departamento de Astronomia, Instituto de Astronomia, Geofísica e Ciências Atmosféricas, Universidade de São Paulo, S\~ao Paulo SP, Brazil}
\begin{document}
%
\date{}
\pagerange{\pageref{firstpage}--\pageref{lastpage}} \pubyear{2012}
\maketitle
\label{firstpage}
%
\begin{abstract}
  We use the stacked gravitational lensing mass profile of four high-mass ($M \gtrsim
  10^{15}M_\odot$) galaxy clusters around $z\approx 0.3$ from Umetsu et al. to fit density profiles
  of phenomenological [Navarro-Frenk-White (NFW), Einasto, Sérsic, Stadel, Baltz-Marshall-Oguri
  (BMO) and Hernquist] and theoretical (non-singular Isothermal Sphere, DARKexp and Kang $\&$ He)
  models of the dark matter distribution. We account for large-scale structure effects, including a
  2-halo term in the analysis. We find that the BMO model provides the best fit to the data as
  measured by the reduced $\chi^2$. It is followed by the Stadel profile, the generalized NFW
  profile with a free inner slope and by the Einasto profile. The NFW model provides the best fit
  if we neglect the 2-halo term, in agreement with results from Umetsu et al. Among the theoretical
  profiles, the DARKexp model with a single form parameter has the best performance, very close to
  that of the BMO profile.  This may indicate a connection between this theoretical model and the
  phenomenology of dark matter halos, shedding light on the dynamical basis of empirical profiles
  which emerge from numerical simulations.

\end{abstract}
%
\begin{keywords}
cosmology: dark matter;  galaxies: clusters; galaxies: halos
\end{keywords}
%
%
\section{Introduction}
%
%
Evidence for the existence of dark matter dates back to \cite{Zwicky_1933} with studies of the
kinematics of galaxies in the Coma Cluster, which required the presence of a massive, smooth and
dark component generating the cluster gravitational potential. More recently, astrophysical and
cosmological observations as well as simulations and theoretical arguments have provided further
indication for the existence of dark matter, and hopes that it may be detected directly in particle
accelerators \citep[see for example][ and references therein]{Frandsen_2012}.  These developments
include the flatness of galaxy rotation curves \citep*[e.g., ][]{Bosma_1978, Bosma_1979,
  Rubin_1978, Rubin_1980}, the mass of galaxy clusters inferred either by their X-ray emission
\cite*[e.g., ][]{Allen_2002} and \cite{Vikhlinin_2006} or by gravitational lensing
\citep[e. g.][]{Clowe_2006, Umetsu_2011_b, Mira_2011}, the acoustic oscillations measured in the
cosmic microwave background \citep[see WMAP papers, e. g][]{WMAP_2011_A} and in galaxy surveys
\citep[e. g.][]{Eisenstein_2005, Sanchez_2012}, and detailed studies of structure formation on
numerical simulations with ever increasing precision \citep[e.g.][]{Springel_2005, Alimi_2012}.

In the standard scenario, dark matter is assumed to be composed of particles that interact
gravitationally but not electromagnetically. Within this picture, simulations of structure
formation have shown a number of interesting results regarding the final states of systems of
gravitationally interacting particles. For example, it has been shown that dark matter is cold,
i.e. its particles must have had non-relativistic velocities around the epoch of recombination,
otherwise structures like galaxies and galaxy clusters would be more diffuse than they appear due
to free-streaming.

Another feature that emerges from simulations, and is confirmed by observations, is that these
systems seem to achieve a final state of equilibrium, displaying nearly {\it universal} density
profiles $\rho\PC{r}$ \citep*{NFW_1996, NFW_1997} and pseudo phase-space profiles
$\rho\PC{r}/\sigma^3\PC{r}$ \citep{Taylor_2001} where $\sigma\PC{r}$ is either the radial or total
velocity dispersion. This is intriguing because these non-collisional systems interact only through
gravitational effects, making them quite different from e.g. a molecular gas in a box. We are then
led to approach the question of how a non-collisional process could bring a gravitational system to
an equilibrium state in a time scale of the age of the Universe.

In an attempt to answer this kind of question, \cite{LyndenBell_1967} developed the mechanism of
violent relaxation, in which the system's constituents, e.g. stars or dark matter particles,
interact mainly with a time-varying average gravitational field, for which the time scale to
achieve an equilibrium state is many orders of magnitude smaller than that of two-particle
interactions. However there are issues in this approach, like infinite masses and mass
segregation. These happen because the model generates density profiles similar to that of an
Isothermal Sphere (see below). For more details on these and other critics to the
\cite{LyndenBell_1967} approach, see e.g. \cite*{Hjorth_Williams_2010_I} and references therein.

Regardless of the discussions about the statistical process involved, the density profiles provided
by simulations are observed in real objects like galaxy clusters, in which the role of the baryonic
component is relatively small compared to that of dark matter. Phenomenologically, one can argue
that if simulations provide density profiles which match those in observed data, that means that
the assumptions made in the simulations are likely correct, and the observed features are
consequences of the gravitational interaction. Nonetheless, a deeper understanding of the physical
mechanisms that lead to equilibrium in gravitational systems is certainly desirable.  In fact,
ignoring this issue would be equivalent to making simulations of a molecular gas in a box and
computing gas pressure and velocity distribution from the simulated results, with no regards to the
kinetic gas theory developed by Maxwell, Boltzmann and others.

In order to have a better dynamical picture of gravitational systems in general, and of dark matter
halos in particular, first principle models have been developed to explain the features seen in
simulations and observations. In particular, there has been a great effort to make predictions of
the three-dimensional density profile $\rho\PC{\bf{r}}$ of dark matter
halos. 
The connection with observations is made via the surface density profile $\Sigma\PC{R}$ projected
in the line-of-sight $x_{\parallel}$
\begin{equation}
\Sigma\PC{R} = \int {dx_{\parallel} \ \rho(x_{\parallel},R}) \,,
\label{eq_sigma_given_rho}
\end{equation}
where ${\bf r}=(x_{\parallel},R)$ and $R$ is the projected distance on the plane of the
sky. Compared to galaxies, for which dissipative effects of cold baryons are important, galaxy
clusters are excellent to test the distribution of dark matter, because in clusters most of the
baryons are hot and dissipate less. Thus, the total density profile, inferred e.g. from
gravitational lensing measurements, provides reliable information about the dark matter density
profile. In fact, lensing is particularly interesting in the determination of the observed density
profile of galaxy clusters, because it does not require assumptions of hydrostatic equilibrium, as
in dynamical methods.

In this paper we use the stacked surface density profile from four massive galaxy clusters with
similar mass and redshift to test both phenomenological and theoretical models for their density
profiles. We only consider spherically symmetrical models. In \S \ref{sec_Data} we briefly describe
the cluster data used in this work.  In \S \ref{sec_modelos_fenom} and \S
\ref{sec_modelos_teoricos} we present the phenomenological and theoretically motivated models
tested. In \S \ref{sec_halomodel} we present the halo model, which allows us to include large-scale
structure effects on the observed profiles.  Our results are presented in \S \ref{sec_results} and
discussed in \S \ref{sec_discuss}. When necessary, we use for the cosmological parameters:
$\Omega_m = 0.275$ and $\Omega_\Lambda = 0.725$.

%

\section{Data}
\label{sec_Data}

We use the data of \cite{Umetsu_2011_b}, who combined weak-lensing shear, magnification, and
strong-lensing measurements of four high-mass ($M \gtrsim 10^{15}M_\odot$) galaxy clusters (A1689,
A1703, A370, C10024+17) with redshifts around $z\approx 0.3$. The strong lensing data was based on
Hubble Space Telescope observations for the central regions of those clusters (typically,
$R~\lesssim~150$ kpc$/h$), and combined with independent weak-lensing data obtained by
\cite{Umetsu_2011_a}, extending to the outer regions ($R~\lesssim~3.5$ Mpc$/h$) of the clusters.

The uncertainties and covariance matrices on the individual surface density profiles were
calculated by \cite{Umetsu_2011_b} taking into account the observational errors as well as ``the
effect of uncorrelated large scale structure projected along the line of sight'', which can be
determined once a cosmological model is assumed. The virial radii and virial masses were obtained
using the non-parametric deprojection method developed by \cite{Broadhurst_Barkana_2008}. A summary
of the properties of the individual clusters is shown in Table~(\ref{tab_properties_clusters})
\citep[data from][]{Umetsu_2011_a}.

\begin{table}
\begin{center}
\begin{tabular}{|l|c|c|c|}
  \hline
  \multicolumn{1}{|c|}{\textbf{Cluster}} & \textbf{Redshift} & \textbf{$r_{vir}$} & \textbf{$M_{vir}$} \\
  &           &  $(Mpc~h^{-1})$ & $(10^{15}M_\odot ~h^{-1})$ \\ \hline
  A1689 & 0.183 & $2.011\pm 0.113$ & $1.300\pm 0.205$ \\ \hline
  A1703 & 0.281 & $1.915\pm 0.148$ & $1.325\pm 0.221$ \\ \hline
  A370 & 0.375 & $2.215\pm 0.079$ & $2.399\pm 0.249$ \\ \hline
  C10024+17 & 0.395 & $1.799\pm 0.105$ & $1.329\pm 0.224$ \\ \hline
\end{tabular}
\end{center}
\caption{Properties of individual clusters: name, redshift, virial radius and virial mass. Data extracted from Tables (1) and (7) of \protect\cite{Umetsu_2011_a}.}
\label{tab_properties_clusters}
\end{table}

In order to have an averaged cluster representative of the sample, also reducing the cosmic noise
and smoothing effects due to asphericity or presence of substructures, \cite{Umetsu_2011_b} built a
stacked surface density profile, scaling the individual cluster profiles by their virial
radii. This procedure is justified given the narrow range in redshift and mass of the individual
clusters [see Table~(\ref{tab_properties_clusters})]. It also takes into account the different
redshifts and their corresponding weight in the calculation of the full covariance matrix
$V_{ij}$.
 

Hereafter we assume that the radial shape of the mean density profile obtained in this way is
representative of dark matter halos in equilibrium \citep{Gao_2012}. For more detailed information
about the observations, uncertainties and stacking process, see \cite{Umetsu_2011_b}.


\section{Phenomenological models}
\label{sec_modelos_fenom}
A number of phenomenological models for density profiles of dark matter halos and galaxy clusters
have been proposed as parametrized functions that fit reasonably well simulations and observations,
with no regards to fundamental principles or theoretical motivation. Below we make a brief
description of the models that we test.

\subsection{NFW profile}
\label{subsec_NFW}
The NFW profile was proposed by \cite*{NFW_1996, NFW_1997} in order to fit the data of N-body cold
dark matter (CDM) simulations, after stacking many halos. It is given by
\begin{equation}
\rho\PC{r} = \frac{\rho_s}{\PC{r/r_s}\PC{ 1 + r/{r_s}}^2 }\,,
\end{equation}
where $\rho_s$ and $r_s$ are scale parameters. It often represents the best fit model to observed data of galaxy clusters; this would also be the case in this work if we did not include large-scale structure effects in the analysis, as discussed below. The NFW profile has an analytical expression for the surface density profile $\Sigma\PC{R}$ \citep{Bartelmann_1996}, given by
\begin{equation}
\Sigma\PC{R} = 2\rho_s r_s F\PC{R/r_s}\,,
\end{equation}
where
\begin{displaymath}
F\PC{X} = \left\{ \begin{array}{ll}
 \cfrac{1}{X^2 - 1}\PC{1 - \cfrac{2}{\sqrt{1 - X^2}}\arctanh\sqrt{\cfrac{1 - X}{1 + X}}}\,, & \textrm{$(X < 1)$}\\
 \cfrac{1}{3}\,, & \textrm{$(X=1)$}\\
 \cfrac{1}{X^2 - 1}\PC{1 - \cfrac{2}{\sqrt{X^2 - 1}}\arctan\sqrt{\cfrac{X - 1}{X + 1}}}\,. & \textrm{$(X > 1)$}
  \end{array} \right.
\end{displaymath}
The NFW profile has a non-physical divergence at the origin, varying as $r^{-1}$ in the inner
regions. In its outer parts it varies as $r^{-3}$, implying another unrealistic property of an
infinite total mass.  One way to circumvent the latter divergence is to truncate the profile at a
maximum radius, e.g. the virial radius.

A common generalization of the NFW profile \citep{Zhao_1996, Jing_Suto_2000} is obtained by setting
the inner slope as a free parameter $\alpha$ (for NFW, $\alpha = -1$):
\begin{equation}
\rho\PC{r} = \frac{\rho_s}{\PC{r/r_s}^\alpha\PC{1 + r/r_s}^{3 - \alpha}}\,,
\end{equation}
Following \cite{Umetsu_2011_b} we will refer to this generalized model as gNFW.

\subsection{BMO profile}
Another useful modification of the NFW profile is that proposed by \cite*{BMO_2009}, which
incorporates a polynomial, smooth, truncation in the outer regions, obtaining a profile steeper
than the NFW in this region. In this way, the problem with the infinite mass of NFW is circumvented
and the influence of the 2-halo term can be better taken into account, as discussed in
\S~\ref{sec_halomodel}. The function proposed has the following shape:
\begin{equation}
\rho\PC{r} = \frac{\rho_s}{\PC{r/r_s}\PC{ 1 + r/{r_s}}^2 }\PC{\frac{r_t^2}{r^2 + r_t^2}}^n\,,
\end{equation}
with $n$ and $r_t$ being free parameters. Actually, we obtained better results in the fits fixing
$n=2$.

\subsection{Sérsic profile}
The Sérsic profile \citep[see ][]{Sersic_1963} was proposed in order to fit the light distribution
in spheroidal galaxies and has been also used to fit simulated data \citep{Merritt_2006}. It is
defined as a projected surface density profile that has the form
\begin{equation}
\Sigma\PC{R} = \Sigma_e \exp\chaves{ {-b_n\PR{\PC{R/R_e}^{1/n} - 1} } }\,,
\end{equation}
where $\Sigma_e$ is the surface brightness at the effective radius $R_e$ and $b_n$ is a function of
$n$ obtained by imposing that the luminosity inside the effective radius is half the total
luminosity. 
The relation between $b_n$ and $n$ is well approximated by $b_n \approx 2n - 0.324$
\citep{Ciotti_1991}.

\subsection{Einasto profile}
The Einasto profile is a three-dimensional version of the Sérsic profile \citep[see
][]{Einasto_1965}. It was proposed to describe the surface brightness of elliptical
galaxies. Recently it has also been used to fit data from N-body CDM simulations, giving results
comparable to the NFW profile in some cases \citep{Navarro_2004, Merritt_2005, Merritt_2006,
  Gao_2008, Navarro_2010}. It is given by
\begin{equation}
\rho\PC{r} = \rho_s \exp\chaves{-2n\PR{\PC{r/r_{-2}}^{1/n} - 1}}\,,
\end{equation}
where $\rho_s$ and $r_{-2}$ are scale parameters.  \cite{Lapi_Cavaliere_2011} discuss a possible
dynamical basis that could generate a profile for which Einasto is a good approximation. In their
Appendix A, \cite*{Mamon_2010} have obtained a polynomial approximation to better than $0.8\%$ for
the expression of the surface density, in the intervals $3.5 \leq n \leq 6.5$ and $-2 \leq
\log_{10}\PC{R/r_{-2}} \leq 2$.

\subsection{Stadel profile}
This profile was proposed to fit simulated data of galaxy-size dark matter halos \citep[see
][]{Stadel_2009}. It has the form
\begin{equation}
\rho\PC{r} = \rho_0 \exp\chaves{-\lambda\PR{\ln\PC{1 + r/r_s}}^2}\,,
\end{equation}
which resembles somewhat the Einasto profile and similarly gives a finite density $\rho_0$ at the
origin. It can also be written as
\begin{equation}
\rho\PC{r} = \frac{\rho_0}{\PC{1 + r/r_s}^{\lambda \ln\PC{1 + r/r_s}}},
\end{equation}
and in this way it resembles power-law profiles. Noticing that the shape parameter assumed almost
the same value $\lambda = 0.1$ in different simulations, \cite{Stadel_2009} proposed to fix this
parameter and promote the model into a two-parameter profile. Here we let $\lambda$ be a free
parameter and obtain a different value for it.

\subsection{Hernquist profile}
The \cite{Hernquist_1990} profile has the functional form
\begin{equation}
\rho\PC{r} = \frac{\rho_s}{\PC{r/r_s}\PC{1 + r/r_s}^3},
\end{equation}
and differs from the NFW profile only in the outer parts, where it varies as $r^{-4}$. It was
proposed, not as a fit to simulated or observed data, but because it provides analytical
expressions for dynamical quantities, such as the gravitational potential, the energy distribution
function, the density of states as well as the surface density, which is given by
\begin{equation}
\Sigma\PC{R} = 2\rho_s r_s G\PC{R/r_s},
\end{equation}
where
\begin{equation}
G\PC{X} =  \frac{\PR{\PC{2 + X^2}H\PC{X} - 3}}{2\PC{1 - X^2}^2},
\end{equation}
and
\begin{displaymath}
H\PC{X} = \left\{
  \begin{array}{ll}
 \cfrac{1}{\sqrt{1 - X^2}}\sech^{-1} X\,, & \textrm{$(X < 1)$}\\
 1\,, & \textrm{$(X = 1)$}\\
 \cfrac{1}{\sqrt{X^2 - 1}}\sec^{-1} X\,, & \textrm{$(X > 1)$}
  \end{array} \right.
\end{displaymath}
which implies that $\displaystyle \lim_{X \to 1} G\PC{X}=  2/15$.

After addition of the 2-halo term explained in \S\ref{sec_halomodel}, the phenomenological profiles
described above are shown in Fig.~\ref{img_best_fit_fenom_corr_2_halos}, along with the galaxy
cluster stacked data from \cite{Umetsu_2011_b}. 


\section{Theoretical models}
\label{sec_modelos_teoricos}
Some of the theoretical models we investigate here are based on the hypothesis of hydrostatic
equilibrium between the gravitational attraction and the pressure $P\PC{r}$ due to velocity
dispersion in an isotropic distribution:
\begin{equation}
\frac{dP}{dr} = -\rho\PC{r}\frac{GM\PC{r}}{r^2},
\label{eq_equil_hidrost_1}
\end{equation}
where $\rho\PC{r}$ is the mass density profile and $M\PC{r}$ is the total mass inside radius $r$:
\begin{equation}
  M\PC{r}=\int_0^r dr'4\pi {r'}^2 \rho \PC{r'} \,.
\label{eq_mass_given_rho}
\end{equation}
Combining Eqs.~(\ref{eq_equil_hidrost_1}) and (\ref{eq_mass_given_rho}) we have
\begin{equation}
\frac{d}{dr}\PR{\frac{r^2}{\rho\PC{r}}\frac{dP}{dr}} = - 4\pi Gr^2\rho\PC{r}.
\label{eq_equil_hidrost_2}
\end{equation}

Choosing the equation of state $P\PC{\rho}$ determines the model, and
Eq.~(\ref{eq_equil_hidrost_2}) can then be (numerically) solved to give the density profile
$\rho\PC{r}$.

\subsection{(non-singular) Isothermal Sphere}
The (non-singular) Isothermal Sphere is based on the equation of state of an ideal gas $P=nk_BT$,
which locally becomes
\begin{equation}
P\PC{r} = \frac{k_BT}{m}\rho\PC{r},
\label{eq_state_eq_ideal_gas}
\end{equation}
where $m$ is the mass of the constituent particle. Using Eq.~(\ref{eq_state_eq_ideal_gas}) in
Eq.~(\ref{eq_equil_hidrost_2}) we have
\begin{equation}
%
%
r\rho\frac{d^2{\rho}}{dr^2} - r{\left(\frac{d\rho}{dr}\right)}^2 + 2\rho\frac{d\rho}{dr} + 4\pi G \lambda r\rho^3 = 0,
\label{eq_diff_Isot_Sphere}
\end{equation}
where $\lambda = m/k_B T$. This represents a particular case of the so-called Lane-Emden
equation. The (non-singular) Isothermal Sphere has null slope at the origin and oscillates around
the Singular Isothermal Sphere $\PC{\rho \propto r^{-2}}$ for large radii \citep{Binney_2008}. Thus
we solve Eq.~(\ref{eq_diff_Isot_Sphere}) numerically imposing the boundary conditions
$d\rho/dr(0)=0$ and $\rho\PC{0}=\rho_0$, where $\rho_0$ is a free parameter.

\subsection{Kang \& He models}
The entropy per unit mass $s_r$ of an ideal gas, written as a function of pressure $p_r$ and
density profile $\rho\PC{r}$ is
\begin{equation}
s_r = \ln\PC{p_r^{3/2}\rho^{-5/2}}\,,
\end{equation}
and the Jeans equation is written as 
\begin{equation}
\frac{dp_r}{dr} + 2\beta\frac{p_r}{r} = -\rho\frac{GM\PC{r}}{r}\,,
\label{eq_Jeans_equation}
\end{equation}
where $\beta = 1 - \PC{\sigma_\theta^2 + \sigma_\phi^2}/\PC{2\, \sigma_r^2}$ is the velocity
anisotropy parameter, written in terms of the velocity dispersions in the three spherical
coordinates. \cite{Kang_He_2011_A} define a generalized pressure $P$ as
\begin{equation}
\frac{dP}{dr} = \frac{dp_r}{dr} + 2\beta\frac{p_r}{r}
\end{equation}
and a phenomenological entropy as
\begin{equation}
s =\ln\PC{P^{3/2}\rho^{-5/2}}\,,
\label{eq_entropy_ideal_gas_generalized}
\end{equation}
such that the resulting system of equations is independent of $\beta$. This effectively reduces
Eq.~(\ref{eq_Jeans_equation}) to Eq.~(\ref{eq_equil_hidrost_1}). Using the variational principle,
the entropy per unit mass, Eq.~(\ref{eq_entropy_ideal_gas_generalized}), is then used to maximize
the total entropy $S$
\begin{equation}
S = \int_0^\infty 4\pi r^2\rho s \ dr = \int_0^\infty 4\pi r^2\rho \ln\PC{P^{3/2}\rho^{-5/2}}dr \,,
\end{equation}
subject to the constraints of conservation of total energy and the virial theorem. This procedure
results in the following equation of state
\begin{equation}
\rho = \lambda P + \mu P^{\gamma}\,,
\label{eq_state_eq_kang_he}
\end{equation}
where $\gamma=3/5$. We will refer to this model as ``Kang $\&$ He''. The constant $\lambda$ is a
Lagrange multiplier and $\mu$ is an integration constant, both related to total mass and energy of
the system. This equation of state reduces to that of an ideal gas
Eq.~(\ref{eq_state_eq_ideal_gas}) for $\mu = 0$ and $\lambda = m/k_BT$. Following a similar but
different approach, \cite{Kang_He_2011_B} obtain the same equation, but now with $\gamma = 4/5$. We
will call this last model ``Kang $\&$ He 2''.

In order to use the equation of state Eq.~(\ref{eq_state_eq_kang_he}) in
Eq.~(\ref{eq_equil_hidrost_2}), we need to solve for $P\PC{\rho}$ in
Eq.~(\ref{eq_state_eq_kang_he}), so as to turn Eq.~(\ref{eq_equil_hidrost_2}) into an equation for
$\rho(r)$. This is not possible for general values of $\gamma$, so after differentiating
Eq.~(\ref{eq_state_eq_kang_he}), \cite{Kang_He_2011_A} propose approximating $P \approx
\rho/\lambda$, obtaining
\begin{equation}
\frac{dP}{dr} = \frac{1}{\lambda + \gamma\mu \PC{\lambda/\rho}^{1 - \gamma}}\frac{d\rho}{dr}\,.
\label{eq_dPdr_KH}
\end{equation}
Inserting Eq.~(\ref{eq_dPdr_KH}) into Eq.~(\ref{eq_equil_hidrost_2}), one obtains a second order
differential equation for $\rho\PC{r}$, which can be numerically solved imposing again
$d\rho/dr(0)=0$ and $\rho(0)=\rho_0$.

It is possible to follow a different approach, inserting $\rho(r)$ from
Eq.~(\ref{eq_state_eq_kang_he}) into Eq.~(\ref{eq_equil_hidrost_2}), thus obtaining an equation for
$P(r)$. After solving this equation numerically for $P(r)$, $\rho(r)$ can be obtained from
Eq.~(\ref{eq_state_eq_kang_he}). This approach proves to give slightly better results (although
similar to the original Kang \& He's ) in the fitting procedure, so that is what we used.

\subsection{DARKexp}\label{subsec_DARKexp}
The DARKexp model \citep*{Hjorth_Williams_2010_I, Hjorth_Williams_2010_II} is significantly
different from the previous models, because it does not take into account a possible equation of
state to be used in the hydrostatic equilibrium, Eq.~(\ref{eq_equil_hidrost_2}). Instead, it deals
with statistical mechanical arguments to (indirectly) derive the distribution function and
determine the density profile.

For the discussion below, let us define a dimensionless density $\tilde{\rho} = {\rho}/{\rho_0}$
and a dimensionless distance $x = {r}/{a}$, where $\rho_0$ and $a$ are scale parameters. The
particle's energy per unit mass $E = \Phi + v^2/2$, where $\Phi$ is the gravitational potential and
$v$ the particle velocity, can be written as
\begin{equation}
\varepsilon = \varphi - \frac{1}{2}\frac{v^2}{{v_g}^2},
\label{eq_energy_dimensionless}
\end{equation}
where $v_g = \sqrt{a^2\rho_0 G}$ and we defined the positive and dimensionless quantities
$\varepsilon = -{E}/{{v_g}^2}$ and $\varphi = -{\Phi}/{{v_g}^2}$.

The DARKexp model is based on two main assumptions. First, because dark matter in halos is
collisionless, it is argued that, after the system reaches an equilibrium, {\it each particle}
retains its individual energy, and thus a Boltzmann-like function must be used, not in the
distribution function $f\PC{\varepsilon}$ (average number of particles per state of energy
$\varepsilon$), but in the number of particles per unit energy $N\PC{\varepsilon} \propto
f\PC{\varepsilon} g\PC{\varepsilon}$, where $g\PC{\varepsilon}$ is the density of states (number of
states per unit energy); see \cite{Binney_1982}. The other feature of the model is that it properly
considers the possibility of low occupation numbers, which results in a cutoff similar to that of
King models \citep{King_1966, Madsen_1996}.  These two features imply that the number of particles
per unit energy $\varepsilon$ must be given by
\begin{equation}
N\PC{\varepsilon} = e^{\varphi_0 -\varepsilon} - 1\,,
\label{eq_DARKexp_model}
\end{equation}
where $\varphi_0$ is the shape parameter representing the central potential.

In models that predict the distribution function $f\PC{\varepsilon}$, the density profile is
obtained after integrating over all possible velocities \citep[see ][]{Binney_2008}
\begin{equation}
\rho\PC{x} = 4\pi\int dv v^2 \frac{\rho_0}{{v_g}^3} f\PC{\varepsilon},
\end{equation}
where $f\PC{\varepsilon}$ is considered to be dimensionless. The equation above also assumes that
the velocities are isotropic, contrary to what is seen in simulated $\Lambda$CDM halos
\citep[e. g.][]{Lemze_2012} and in observational analysis \citep[e. g.][]{Biviano_2004}. With the
help of Eq.~(\ref{eq_energy_dimensionless}), we have
\begin{equation}
  \tilde{\rho}\PC{x} = 4\pi \int_0^{\varphi\PC{x}} d\varepsilon f\PC{\varepsilon}\sqrt{2\PR{\varphi\PC{x} - \varepsilon}}\,.
\label{eq_rho_given_f}
\end{equation}
The density profile is finally obtained by solving Poisson's equation
\begin{equation}
\nabla^2\varphi\PC{x} = -4\pi \tilde{\rho}\PC{x}\,.
\end{equation}

However, if the model predicts $N\PC{\varepsilon}$, as in the case of the DARKexp, we need to use
an iterative approach. Here we follow the procedure of \cite{Binney_1982}. We start by guessing an
initial estimate of the density profile $\tilde{\rho}\PC{x}$ and calculate the resulting potential
as
\begin{equation}
\varphi\PC{x} = 4\pi \PR{\frac{1}{x}\int_0^{x}dx' x'^2\tilde{\rho}\PC{x'} +\int_{x}^{\infty}dx' x'\tilde{\rho}\PC{x'}}\,.
\label{eq_psi_given_rho}
\end{equation}
Next, we compute the density of states as
\begin{equation}
g\PC{E} = \PC{4\pi}^2\int dr r^2 \int dv v^2 \delta\PC{\frac{1}{2}v^2 + \Phi - E}\,
\end{equation}
which in terms of the dimensionless quantities results in 
\begin{equation}
  g\PC{\varepsilon} = 16\pi^2 a^3 v_g \int_0^{x_{max}\PC{\varepsilon}} dxx^2\sqrt{2\PR{\varphi\PC{x} - \varepsilon}}\,,
\end{equation}
where $x_{max}$ is such that $\varphi\PC{x_{max}} = \varepsilon$. We then use the
$N\PC{\varepsilon}$ defined in the model, Eq.~(\ref{eq_DARKexp_model}), to compute the
dimensionless distribution function
\begin{equation}
f\PC{\varepsilon} = a^3 v_g \frac{N\PC{\varepsilon}}{g\PC{\varepsilon}}\,.
\end{equation}
Finally, we use Eq.~(\ref{eq_rho_given_f}) to obtain a new $\tilde{\rho}\PC{x}$ and iterate the
process. We find that after about 20 iterations the model converges to a density profile
independent of the initial guess.

Fig.~\ref{img_best_fit_teor_corr_2_halos} shows the theoretical models described above, after
adding to them the 2-halo term explained in \S\ref{sec_halomodel}, along with the galaxy cluster
stacked data from \cite{Umetsu_2011_b}.


\section{Halo Model}
\label{sec_halomodel}

When considering cluster profiles that extend to sufficiently large radii, large-scale corrections
must be taken into account. For dark matter halos of a given mass $M$ and redshift $z$, the
halo-mass correlation function, defined as $\xi_{hm}(r)=\langle \delta_h({\bf x}) \delta_m({\bf x +
  r}) \rangle$, represents the excess density of matter at a distance $r=|{\bf r}|$ from the halo
center, i.e. it is a measure of the average observed halo profile $\langle \rho_{\rm
  obs}(r)\rangle$: \bea 1+ \xi_{hm}(r)=\frac{\langle \rho_{\rm obs}(r) \rangle}{\bar{\rho}_m} \,.
\label{eq:xihm_def}
\eea

The halo model \citep[see][]{Cooray_2002} allows us to estimate cosmological correlations from the
properties of dark matter halos, seen as the building blocks of cosmic structure. In this context
the halo-mass correlation function is given by a sum of two contributions \citep[][]{Hayashi_2008,
  Schmidt_2009} \bea \xi_{hm}(r)= \frac{\rho_{1h}(r)}{\bar{\rho}_m} + b_h^{L}(M)\xi_{m}^L(r)\,.
\label{eq:xihm_halomodel}
\eea

Here $\rho_{1h}(r)$ represents the 1-halo contribution or true halo profile from matter within the
halo itself; this is the term described by all models presented in \S\ref{sec_modelos_fenom} and
\S\ref{sec_modelos_teoricos}.  The second term on the right-hand side of
Eq.~(\ref{eq:xihm_halomodel}) represents the 2-halo contribution from the large-scale structure of
the Universe, given by the linear matter correlation function $\xi_{m}^L(r)$ and the linear halo
bias $b_h^L(M)$.

Projected lensing measurements are sensitive to the average observed overdensity $\delta \rho_{\rm
  obs}(r) = \langle \rho_{\rm obs}(r) \rangle - \bar{\rho}_m$. Therefore, combining
Eqs.~(\ref{eq:xihm_def}) and (\ref{eq:xihm_halomodel}) we find \bea \delta \rho_{\rm obs}(r) =
\rho_{1h}(r) + \rho_{2h}(r) \,, \eea where the 2-halo term is given by \bea
\rho_{2h}(r)=\bar{\rho}_m b_h^L(M)\xi_m^{L}(r)\,.
\label{eq_2halo_term}
\eea

The observed surface density profile $\Sigma_{\rm obs}(R)$ at projected distance $R$ is obtained
using Eq.~(\ref{eq_sigma_given_rho}): \bea \Sigma_{\rm obs}(R)=\int dx_{\parallel} \ \delta
\rho_{\rm obs}(x_{\parallel},R) 
=\Sigma_{1h}(R)+\Sigma_{2h}(R)\,, \eea where $\Sigma_{1h}(R)$ is defined from $\rho_{1h}(r)$, and
similarly for $\Sigma_{2h}(R)$. We estimate $b^L_h(M)$ from the fit to simulations of
\cite{Tinker_2010} and $\xi_m^L(r)$ as the Fourier transform of the linear matter power spectrum
$P_m^L(k)$ obtained from CAMB \citep*{Lewis_2000}, \bea \xi_m^L(r)=\frac{1}{2\pi^2}\int dk \ k^2
P_m^L(k) \frac{\sin(kr)}{kr}\,.  \eea Finally we assume a flat Universe with cosmological
parameters $\Omega_m = 0.275$ and $\Omega_\Lambda = 0.725$.

\begin{figure}
\begin{center}
\includegraphics[scale=0.42]{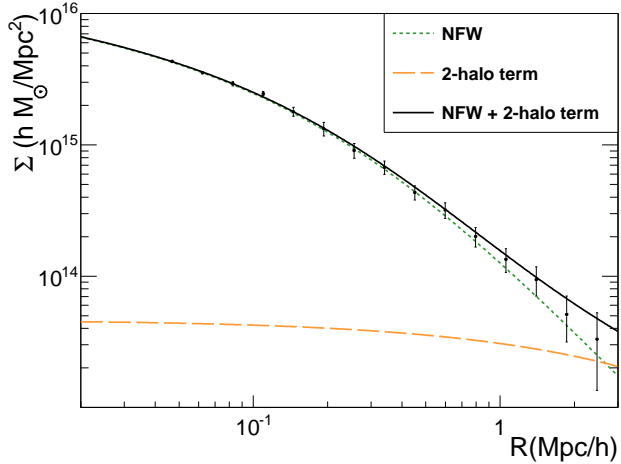}
\caption{NFW profile (dotted), 2-halo term correction (dashed) and the sum of both (solid)
  representing the final best fit are shown, along with data points from
  \protect\cite{Umetsu_2011_b}.}
\label{img_best_fit_NFW_corr_2_halos}
\end{center}
\end{figure}

In Fig.~\ref{img_best_fit_NFW_corr_2_halos} we illustrate the effect of the 2-halo term for massive
halos. The average redshift of the sample is $z=0.32$ [see
  Table~(\ref{tab_properties_clusters})]. For this redshift and cosmology, the virial overdensity
  relative to the mean matter density is $\Delta \approx 263$ \citep{Bryan_Norman_1998}. As a first
  approximation, we fit a NFW profile to the stacked halo data, obtaining a mass $M_{vir}=1.56
  \times 10^{15}M_{\odot}/h$. With this mass and redshift, the \cite{Tinker_2010} fitting formula
  gives $b_L(M_{263},z)=10.98$ for the bias factor, and after calculating the linear matter
  correlation function we finally obtain the 2-halo term Eq.~(\ref{eq_2halo_term}). This is an
  iterative process, in which the new mass obtained could be used to calculate a new bias factor
  until it converges, but in this work we restricted the calculation to this first order
  correction.

  Fig.~\ref{img_best_fit_NFW_corr_2_halos} shows that the NFW model underestimates the observed
  profile for $r>1$ Mpc/$h$, where the 2-halo term becomes increasingly important. We add the
  computed $\Sigma_{2h}$ to the 1-halo models of \S\ref{sec_modelos_fenom} and
  \S\ref{sec_modelos_teoricos} before fitting them to data. Since the 2-halo contribution depends
  only on the fixed cosmology, this does not introduce any extra parameter.

\section{Results}
\label{sec_results}
We fit the various models to the data with the help of the Minuit package developed by
\cite{Minuit}. We compute the $\chi^2$, i.e. the minimum value of $Q$, given by
\begin{equation}
Q = \Delta_i \ V_{ij}^{-1} \Delta_j \,,
\end{equation} 
where
\begin{equation}
\Delta_i = \Sigma_{T}\PC{R_i} - \Sigma_{D}\PC{R_i}\,,
\end{equation} 
$\Sigma_{T}\PC{R_i}$ is the surface density from Eq.~(\ref{eq_sigma_given_rho}) for a given model
evaluated at radius $R_i$, $\Sigma_{D}\PC{R_i}$ is the surface mass density from
\cite{Umetsu_2011_b} lensing analysis of the data,
and $V_{ij}$ is the error covariance matrix between data points $i$ and $j$ (see
\S~\ref{sec_Data}). The data consist of 15 correlated points.

Hereafter, the metric we use to compare the various models is the $\chi^2$ per degree of freedom,
or reduced $\chi^2$, defined as
\begin{equation}
\chi_\nu^2 = {\chi^2\over \nu}\,
\label{eq_chi_2_red}
\end{equation}
where $\nu = 15 - N_p$ is the number of degrees of freedom given 15 data points and $N_p$
parameters.

We note that a more rigorous statistical analysis would involve the correct calculation of the
integral of the $\chi^2$ distribution. Such detailed analysis is beyond the scope of this work,
since we are mainly interested in investigating what class of models provide a reasonable
description of the data, as inferred from a simple ranking criterion. We believe though that our
conclusions would remain unchanged if a more detailed statistical analysis was employed.

\begin{figure}
\begin{center}
\includegraphics[scale=0.42]{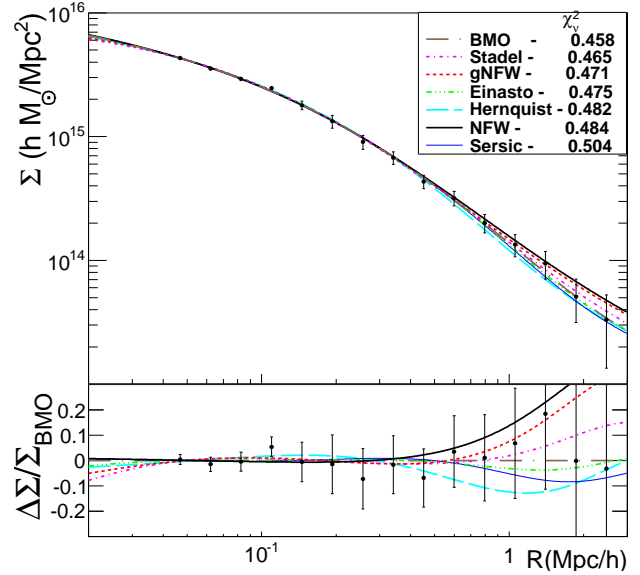}
\caption{Best fit curves and $\chi^2_\nu$ for the phenomenological models along with data points
  from \protect\cite{Umetsu_2011_b}. The bottom panel shows the relative difference between the
  best fits of each model and the BMO profile.}
\label{img_best_fit_fenom_corr_2_halos}
\end{center}
\end{figure}

Fig.~\ref{img_best_fit_fenom_corr_2_halos} shows the data points obtained by \cite{Umetsu_2011_b}
and the best fits for all the phenomenological models discussed above, after the addition of the
2-halo term represented by the orange dashed line in Fig.~\ref{img_best_fit_NFW_corr_2_halos}.

The BMO profile, with 3 parameters and $\chi^2_\nu = 0.458$, represents the best fit, with best fit
value of the truncation radius given by $\tau = r_t/r_s = 17 \pm 15$. The Stadel profile, with 3
parameters and $\chi^2_\nu = 0.465$, represents the second best fit, with best value of the shape
parameter $\lambda = 0.25 \pm 0.04$. It is followed by the generalized version of NFW, gNFW (3
parameters), with $\chi^2_\nu = 0.471$ and $\alpha = 0.74 \pm 0.44$. Next is the Einasto profile (3
parameters), with $\chi^2_\nu = 0.475$ and $n = 3.80\pm 0.61$. Then we have the Hernquist model (2
scale parameters) with $\chi^2_\nu = 0.482$, followed by the NFW model (2 scale parameters) with
$\chi^2_\nu = 0.484$. Finally, for the S\'ersic profile we obtain $\chi^2_\nu = 0.504$ and $n =
2.42 \pm 0.34$. These results are summarized in Table~\ref{tab_best_fit_fenom_corr_2_halos}.

\begin{table}
\begin{center}
\begin{tabular}{|l|c|c|c|}
\hline
\multicolumn{1}{|c|}{\textbf{Profile}} & \textbf{$N_p$} & \textbf{$\chi^2_\nu$} & \textbf{Shape parameter} \\ \hline
BMO & 3 & 0.458 & $\tau = 17 \pm 15$ \\ \hline
Stadel & 3 & 0.465 & $\lambda = 0.25 \pm 0.04$ \\ \hline
gNFW & 3 & 0.471 & $\alpha = 0.74\pm 0.44$ \\ \hline
Einasto & 3 & 0.475 & $n = 3.80\pm 0.61$ \\ \hline
Hernquist & 2 & 0.482 & - \\ \hline
NFW & 2 & 0.484 & - \\ \hline
Sérsic & 3 & 0.504 & $n = 2.42 \pm 0.34$\\ \hline
\end{tabular}
\end{center}
\caption{Fit results for the phenomenological models. The column $N_p$ indicates the total number of model parameters, $\chi^2_\nu$ shows the reduced $\chi^2$ defined in Eq.~(\ref{eq_chi_2_red}) and the last column shows the best estimate for the shape parameter of the model.}
\label{tab_best_fit_fenom_corr_2_halos}
\end{table}

\begin{figure}
\begin{center}
\includegraphics[scale=0.42]{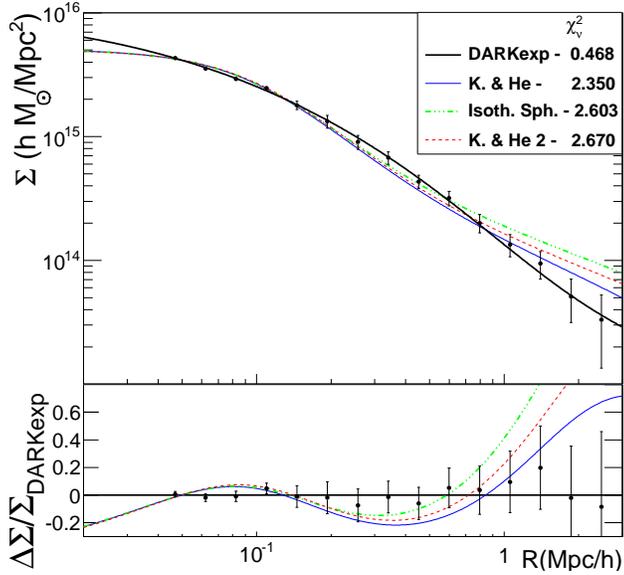}
\caption{Best fit curves and $\chi^2_\nu$ for the theoretical models, along with data points from
  \protect\cite{Umetsu_2011_b}. The bottom panel shows the relative difference between the best
  fits of each model and DARKexp.}
\label{img_best_fit_teor_corr_2_halos}
\end{center}
\end{figure}

Fig.~\ref{img_best_fit_teor_corr_2_halos} shows the fits for the theoretical models. The best fit
was obtained for the DARKexp model with $\chi^2_\nu = 0.468$. In order to generate this model, we
did 25 iterations of the procedure described in \S~\ref{subsec_DARKexp}, with $10^5$ logarithmic
bins in $r$. The best fit value for the shape parameter was $\varphi_0 = 3.00\pm 0.48$. The other
three models, the Isothermal Sphere and the 2 variants predicted by Kang \& He, were generated in
$10^6$ logarithmic bins in $r$.

Among these last three models, the best fit is for the ``Kang $\&$ He'' model, with $\gamma = 3/5$,
for which $\chi^2_\nu = 2.350$ with $\tilde{\lambda} =\lambda c^2 = \PC{5.44\pm 0.10}\times 10^4$
and $\tilde{\mu}= \mu\PC{{c^3}/{\rho_0}}^{2/5} = 13.85\pm 0.03$ for the shape parameters, where $c$
is the speed of light in vacuum and $\rho_0$ is the scale parameter for the density. The Isothermal
Sphere gives $\chi^2_\nu = 2.603$ and $\tilde{\lambda}=\PC{5.76\pm 0.16}\times 10^4$ for the shape
parameter, followed by ``Kang $\&$ He 2'', with $\gamma=4/5$, for which $\chi^2_\nu = 2.670$ with $
\tilde{\lambda}=\PC{5.29\pm 0.38}\times10^4$ and $\tilde{\mu} = \mu\PC{{c^8}/{\rho_0}}^{1/5}= 346
\pm 268$. These results are summarized in Table~\ref{tab_best_fit_teoricos_corr_2_halos}.

\begin{table}
\begin{center}
\begin{tabular}{|l|c|c|c|}
  \hline
  \multicolumn{1}{|c|}{\textbf{Profile}} & \textbf{$N_p$} & \textbf{$\chi^2_\nu$} & \textbf{Shape parameter} \\ \hline
  DARKexp & 3 & 0.468 & $\varphi_0=3.00\pm 0.48$ \\ \hline
  KH & 4 & 2.350 & $\tilde{\lambda} =\PC{5.44\pm 0.10}\times 10^4$\\
  &     &           & $\tilde{\mu}=13.85\pm 0.03$ \\ \hline
  Isoth. Sph. & 3 & 2.603 & $\tilde{\lambda}=\PC{5.76\pm 0.16}\times 10^4$ \\ \hline
  KH2 & 4 & 2.670 & $\tilde{\lambda}=\PC{5.29\pm 0.38}\times10^4$\\
  &    &           & $\tilde{\mu}= 346 \pm 268$ \\ \hline
\end{tabular}
\end{center}
\caption{Fit results for the theoretical models. Columns defined as in Table~\protect\ref{tab_best_fit_fenom_corr_2_halos}.}
\label{tab_best_fit_teoricos_corr_2_halos}
\end{table}

\subsection{Neglecting the 2-halo term}
\begin{figure}
\begin{center}
\includegraphics[scale=0.42]{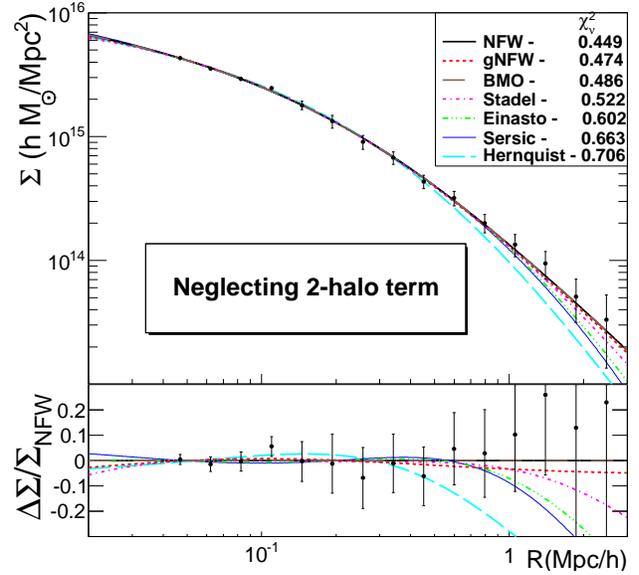}
\caption{Best fit curves and $\chi^2_\nu$ for the phenomenological models studied, neglecting the
  2-halo term, along with data points from \protect\cite{Umetsu_2011_b}. The bottom panel shows the
  relative difference between the best fits of each model and NFW.}
\label{img_best_fit_fenom}
\end{center}
\end{figure}

\begin{figure}
\begin{center}
\includegraphics[scale=0.42]{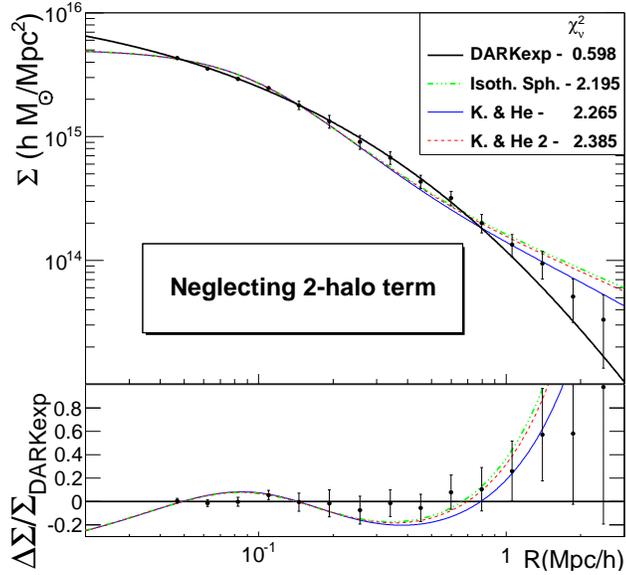}
\caption{Best fit curves and $\chi^2_\nu$ for the theoretical models studied, neglecting the 2-halo
  term, along with data points from \protect\cite{Umetsu_2011_b}. The bottom panel shows the
  relative difference between the best fits of each model and DARKexp model.}
\label{img_best_fit_teoricos}
\end{center}
\end{figure}

We have also considered the results of fitting the models without adding the 2-halo term. These
fits are summarized in Figs.~\ref{img_best_fit_fenom} and \ref{img_best_fit_teoricos} and
Tables~\ref{tab_best_fit_fenom} and \ref{tab_best_fit_teoricos} for the phenomenological and
theoretical models respectively.

In this scenario, the NFW profile produces the overall best fit, with $\chi_\nu^2 = 0.449$ and the
gNFW profile results in $\chi^2_\nu = 0.474$, with the best fit value $\alpha =0.89\pm 0.37$. These
values are identical to those obtained by \cite{Umetsu_2011_b}, who did not include the 2-halo term
in their analysis, and provides a consistency check of our numerical scheme. The third best
  fit is the BMO profile with $\chi^2_\nu = 0.486$ and $\tau = (3\pm 124) \times 10^3$ (what in
  practice makes the model identical to the NFW and shows the complete inadequacy of adding this
  parameter in this case), followed by Stadel profile with $\chi^2_\nu = 0.522$ and $\lambda =
0.223\pm 0.040$, and by the Einasto profile with $\chi^2_\nu = 0.602$ and $n = 4.31 \pm
0.75$. Next, the Sérsic profile resulted in $\chi^2_\nu = 0.663$ with $n = 2.69\pm 0.41$. Finally,
for the Hernquist profile, with 2 scale parameters like NFW, we obtained $\chi^2_\nu =
0.706$. These results are summarized in Table~\ref{tab_best_fit_fenom}.

\begin{table}
\begin{center}
\begin{tabular}{|l|c|c|c|}
  \hline
  \multicolumn{1}{|c|}{\textbf{Profile}} & \textbf{$N_p$} & \textbf{$\chi^2_\nu$} & \textbf{Shape parameter} \\ \hline
  NFW & 2 & 0.449 & - \\ \hline
  gNFW & 3 & 0.474 & $\alpha = 0.89\pm 0.37$ \\ \hline
  BMO & 3 & 0.486 & $\tau = (3 \pm 124) \times 10^3$ \\ \hline
  Stadel & 3 & 0.522 & $\lambda = 0.223 \pm 0.04$ \\ \hline
  Einasto & 3 & 0.602 & $n = 4.31\pm 0.75$ \\ \hline
  Sérsic & 3 & 0.663 & $n = 2.69 \pm 0.41$\\ \hline
  Hernquist & 2 & 0.706 & - \\ \hline
\end{tabular}
\end{center}
\caption{Fit results for the phenomenological models neglecting the 2-halo term. Columns defined as in Table~\protect\ref{tab_best_fit_fenom_corr_2_halos}.}
\label{tab_best_fit_fenom}
\end{table}

Fig.~\ref{img_best_fit_teoricos} shows the fits for the theoretical models when we neglect the
2-halo term. The best fit comes from the DARKexp model, for which $\chi^2_\nu = 0.598$. The best
fit value for the shape parameter was $\varphi_0 = 3.24\pm 0.48$.

Among the Isothermal Sphere and its variants, the best fit is for the former, which provides
$\chi^2_\nu = 2.195$ with $\tilde{\lambda}=\PC{5.62\pm 0.15}\times 10^4$. For the "Kang \& He"
model, with $\gamma = 3/5$, we find $\chi^2_\nu = 2.265 $ with $\tilde{\lambda} = \PC{5.477\pm
  0.003}\times 10^4$ and $\tilde{\mu} = 6.1\pm 0.6$. Finally, for the ``Kang \& He 2'' model with
$\gamma=4/5$, we obtain $\chi^2_\nu = 2.385$ with $\tilde{\lambda} = \PC{5.52\pm 0.36}\times10^4$
and $\tilde{\mu} = 75 \pm 237$. The results of the fits of these theoretical models are summarized
in Table~\ref{tab_best_fit_teoricos}.

As can be seen in Figs.~\ref{img_best_fit_fenom} and \ref{img_best_fit_teoricos}, almost all the
best fit density profiles remain below the data points in the outer regions. The exceptions are the
Isothermal Sphere variants, whose fits nonetheless fail badly. This trend shows the need for
including the 2-halo term in the analysis.

\begin{table}
\begin{center}
\begin{tabular}{|l|c|c|c|}
  \hline
  \multicolumn{1}{|c|}{\textbf{Profile}} & \textbf{$N_p$} & \textbf{$\chi^2_\nu$} & \textbf{Shape parameter} \\ \hline
  DARKexp & 3 & 0.598 & $\varphi_0=3.24\pm 0.48$ \\ \hline
  Isoth. Sph. & 3 & 2.195 & $\tilde{\lambda}=\PC{5.62\pm 0.15}\times 10^4$ \\ \hline
  KH & 4 & 2.265  & $\tilde{\lambda} = \PC{5.477\pm 0.003}\times 10^4$\\
  &     &           & $\tilde{\mu}=6.1\pm 0.6$ \\ \hline
  KH2 & 4 &  2.385 & $\tilde{\lambda} = \PC{5.52\pm 0.36}\times10^4$\\
  &    &           & $\tilde{\mu} = 75\pm237$ \\ \hline
\end{tabular}
\end{center}
\caption{Fit results for the theoretical models neglecting the 2-halo term. Columns defined as in Table~\protect\ref{tab_best_fit_fenom_corr_2_halos}. }
\label{tab_best_fit_teoricos}
\end{table}


\section{Discussion}
\label{sec_discuss}

We have used observed data for the surface mass density of 4 clusters of similar mass and redshift
to study various models for cluster density profiles. Under the assumption that the stacked data
provide a fair representation of the mean radial density profile of dark matter halos, we can
include effects of large-scale structure at large radii and investigate how appropriately each
model describes the average properties of clusters of this mass and redshift, and directly compare
models against each other.
 
For the phenomenological profiles, we have found that the BMO model provides the best fit, followed
by Stadel, gNFW and Einasto. Nonetheless, the performance of all these profiles, including the
standard NFW, is very similar, as is often the case in numerical simulations \citep{Gao_2008,
  Merritt_2005, Navarro_2010}.

For the theoretically motivated profiles, both the Isothermal Sphere and the Kang \& He models give
poor results compared to the phenomenological models. This can be attributed in part to the fact
that both models produce cored density profiles, while the simulations and the data used here favor
cuspy profiles, or cored just in the innermost, inaccessible region, like for the Einasto and
Stadel profiles. Moreover, the outer region is not well described by these models, which behave
like $r^{-2}$, while the data favor a behavior closer to $r^{-3}$.

\begin{figure}
\begin{center}
\includegraphics[scale=0.42]{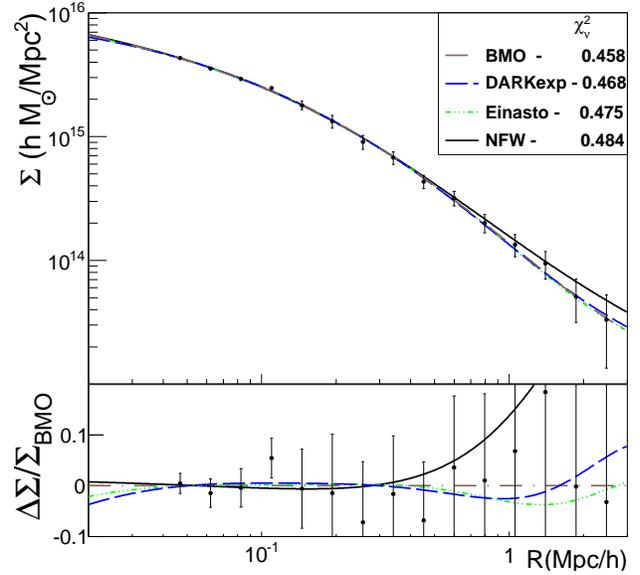}
\caption{Best fit curves for BMO, DARKexp, Einasto and NFW models, along with data points from
  \protect\cite{Umetsu_2011_b}. The bottom panel shows differences relative to the BMO profile.}
\label{img_best_fit_DARKexp_Einasto_NFW}
\end{center}
\end{figure}

The best theoretical fit to data is obtained with the DARKexp model. This model provides an
excellent fit, even compared with the performance of the phenomenological profiles. This is
interesting, since this model has a dynamical basis justification.  We did not investigate the role
of the velocity anisotropy in this model. However, \cite*{Hjorth_Williams_2010_III} have compared
the model to simulated data and shown that the typical anisotropy profiles do not alter
significantly the predicted density profile and that the DARKexp model is a better match to the
Einasto profile than to the NFW profile. This is in agreement with our findings as seen in
Fig.~\ref{img_best_fit_DARKexp_Einasto_NFW}, which shows our best fits for the BMO, DARKexp,
Einasto and NFW profiles.

It is interesting to note that the Einasto profile with $n\approx 6$ generally fits $\Lambda$CDM
simulations better than NFW, as has been noticed by \citet{Navarro_2004}. Moreover
\cite*{Mamon_2010} found $n \approx 5$ for dark matter halos of hydrodynamical cosmological
simulations, while in this work we have found $n \approx 3.8$. Finally, by fitting rotation curves
of spiral galaxies, \citet*{Chemin_2011} have found best fits for the Einasto profile with $n$ as
low or lower than unity. This sequence may suggest a correlation between the parameter $n$ and the
importance of the gas component in the dynamics of the system.

Since the uncertainties, mainly in the external region of the clusters, do not allow us to
statistically discriminate between the phenomenological and DARKexp models, we can conclude that
the latter represents the data as well as the former but the cored profiles (Isothermal Sphere and
its variants) cannot do so.

Regarding the halo model, some cautionary remarks about the methods employed here should be
made. We do not expect the simplified halo model considered to describe accurately the transition
between 1 and 2-halo terms. It is possible to improve on this prescription by including non-linear
effects around the virial radius, e.g. by including the non-linear power spectrum and the
non-linear cluster bias perturbatively \cite{Baldauf_2010}. Such a detailed analysis is important
to accurately characterize the differences between models in more complete cluster samples, but it
is beyond the scope of this work.

In order to roughly access some of the potential degeneracies brought in by non-linear effects, we
allow for the linear bias in the 2-halo term to be a free parameter, instead of fixing it to the
value from the \cite{Tinker_2010} formula.  In this case we obtain the constrains shown in Table
(\ref{tab_best_fit_fenom_free_bias}), and the best-fit value for the bias indicates a bit smaller
average mass $M_{vir}=1.52 \times 10^{15}M_{\odot}/h$ for the stacked clusters, again as calculated
with the NFW density profile. These results indicate that our best-fits are degenerate with the
cluster bias, and likely with parameters describing non-linearities beyond the 1-halo
term. Characterizing these non-linear effects precisely and breaking the extra degeneracies induced
by them would require a more numerous sample of clusters spanning a larger range in mass and
redshift.

\begin{table}
\begin{center}
\begin{tabular}{|l|c|c|c|}
  \hline
  \multicolumn{1}{|c|}{\textbf{Profile}} & \textbf{$N_p$} & \textbf{$\chi^2_\nu$} & \textbf{Bias} \\ \hline
  NFW & 3 & 0.472 & $3.75 \pm 9.14$\\ \hline
  gNFW & 4 & 0.486 & $ 5.71 \pm 9.38$ \\ \hline
  Hernquist & 3 & 0.498 & $15.65 \pm 8.75$ \\ \hline
  BMO & 4 & 0.497 & $13.64 \pm 13.10$ \\ \hline
  Stadel & 4 & 0.505 & $9.45 \pm 11.0$ \\ \hline
  Einasto & 4 & 0.511 & $13.95\pm 10.54$ \\ \hline
  Sérsic & 4 & 0.530 & $15.73 \pm 10.29$\\ \hline
\end{tabular}
\end{center}
\caption{Fit results for the phenomenological models when the bias is a free parameter. The column $N_p$ indicates the total number of model parameters, $\chi^2_\nu$ shows the reduced $\chi^2$ defined in Eq.~(\ref{eq_chi_2_red}) and the last column shows the best estimate for the bias.}
\label{tab_best_fit_fenom_free_bias}
\end{table}

Concluding, we find that, while the observed data is best fit by phenomenological models, there is
a similarly good fit by the theoretical DARKexp model. We find that including the 2-halo term in
the analysis is important and the best-fit ranking somewhat change if it is neglected. Should the
agreement with observational data hold for clusters observed at different ranges of mass and
redshift, one could argue that theoretical models such as DARKexp may provide a dynamical basis for
the observed dark matter density profiles.

\section*{Acknowledgments}
We thank Keiichi Umetsu for providing the cluster data and useful comments, Gary Mamon for critical
reading of the manuscript and suggestions, Dong-Biao Kang for discussions about features of his
models and Liliya L. R. Williams for discussions that helped us reproduce the predictions of the
DARKexp model. This work 
is supported by CNPq, CAPES and FAPESP agencies.
%

%
\bibliographystyle{mn2e}
\bibliography{../../references_leandro}

\begin{thebibliography}{}

\bibitem[\protect\citeauthoryear{Alimi, Bouillot, Rasera, Reverdy, Corasaniti,
  Balmes, Requena, Delaruelle \& Richet}{Alimi et~al.}{2012}]{Alimi_2012}
Alimi J.-M.,  Bouillot V.,  Rasera Y.,  Reverdy V.,  Corasaniti P.-S.,  Balmes
  I.,  Requena S.,  Delaruelle X.,    Richet J.-N.,  2012, arXiv:1206.2838

\bibitem[\protect\citeauthoryear{Allen, Schmidt \& Fabian}{Allen
  et~al.}{2002}]{Allen_2002}
Allen S.~W.,  Schmidt R.~W.,    Fabian A.~C.,  2002, MNRAS, 334, L11

\bibitem[\protect\citeauthoryear{{Baldauf}, {Smith}, {Seljak} \&
  {Mandelbaum}}{{Baldauf} et~al.}{2010}]{Baldauf_2010}
{Baldauf} T.,  {Smith} R.~E.,  {Seljak} U.,    {Mandelbaum} R.,  2010, PRD, 81,
  063531

\bibitem[\protect\citeauthoryear{{Baltz}, {Marshall} \& {Oguri}}{{Baltz}
  et~al.}{2009}]{BMO_2009}
{Baltz} E.~A.,  {Marshall} P.,    {Oguri} M.,  2009, JCAP, 1, 15

\bibitem[\protect\citeauthoryear{Bartelmann}{Bartelmann}{1996}]{Bartelmann_1996}
Bartelmann M.,  1996, A\&A, 313, 697

\bibitem[\protect\citeauthoryear{{Binney}}{{Binney}}{1982}]{Binney_1982}
{Binney} J.,  1982, MNRAS, 200, 951

\bibitem[\protect\citeauthoryear{Binney \& Tremaine}{Binney \&
  Tremaine}{2008}]{Binney_2008}
Binney J.,  Tremaine S.,  2008, Galactic Dynamics - Second Edition.
Princeton University Press

\bibitem[\protect\citeauthoryear{{Biviano} \& {Katgert}}{{Biviano} \&
  {Katgert}}{2004}]{Biviano_2004}
{Biviano} A.,  {Katgert} P.,  2004, AAp, 424, 779

\bibitem[\protect\citeauthoryear{Bosma}{Bosma}{1978}]{Bosma_1978}
Bosma A.,  1978, PhD thesis, Groningen University

\bibitem[\protect\citeauthoryear{Bosma \& van~der Kruit}{Bosma \& van~der
  Kruit}{1979}]{Bosma_1979}
Bosma A.,  van~der Kruit P.~C.,  1979, A\&A, 79, 281

\bibitem[\protect\citeauthoryear{Broadhurst \& Barkana}{Broadhurst \&
  Barkana}{2008}]{Broadhurst_Barkana_2008}
Broadhurst T.~J.,  Barkana R.,  2008, MNRAS, 390, 1647

\bibitem[\protect\citeauthoryear{Bryan \& Norman}{Bryan \&
  Norman}{1998}]{Bryan_Norman_1998}
Bryan G.~L.,  Norman M.~L.,  1998, ApJ, 495, 80

\bibitem[\protect\citeauthoryear{{Chemin}, {de Blok} \& {Mamon}}{{Chemin}
  et~al.}{2011}]{Chemin_2011}
{Chemin} L.,  {de Blok} W.~J.~G.,    {Mamon} G.~A.,  2011, AJ, 142, 109

\bibitem[\protect\citeauthoryear{Ciotti}{Ciotti}{1991}]{Ciotti_1991}
Ciotti L.,  1991, A\&A, 249, 99

\bibitem[\protect\citeauthoryear{Clowe, Bradač, Gonzalez, Markevitch, Randall,
  Jones \& Zaritsky}{Clowe et~al.}{2006}]{Clowe_2006}
Clowe D.,  Bradač M.,  Gonzalez A.~H.,  Markevitch M.,  Randall S.~W.,  Jones
  C.,    Zaritsky D.,  2006, ApJ, 648, L109

\bibitem[\protect\citeauthoryear{Cooray \& Sheth}{Cooray \&
  Sheth}{2002}]{Cooray_2002}
Cooray A.,  Sheth R.,  2002, Physics Reports, 372, 1

\bibitem[\protect\citeauthoryear{{Einasto}}{{Einasto}}{1965}]{Einasto_1965}
{Einasto} J.,  1965, Trudy Astrofizicheskogo Instituta Alma-Ata, 5, 87

\bibitem[\protect\citeauthoryear{Eisenstein, Zehavi, Hogg, Scoccimarro,
  Blanton, Nichol, Scranton, Seo, Tegmark, Zheng, Anderson, Annis, Bahcall,
  Brinkmann, Burles, Castander, Connolly \& et al.}{Eisenstein
  et~al.}{2005}]{Eisenstein_2005}
Eisenstein D.~J.,  Zehavi I.,  Hogg D.~W.,  Scoccimarro R.,  Blanton M.~R.,
  Nichol R.~C.,  Scranton R.,  Seo H.-J.,  Tegmark M.,  Zheng Z.,  Anderson
  S.~F.,  Annis J.,  Bahcall N.,  Brinkmann J.,  Burles S.,  Castander F.~J.,
  Connolly A.,    et al. 2005, ApJ, 633, 560

\bibitem[\protect\citeauthoryear{Frandsen, Kahlhoefer, Preston, Sarkar \&
  Schmidt-Hoberg}{Frandsen et~al.}{2012}]{Frandsen_2012}
Frandsen M.~T.,  Kahlhoefer F.,  Preston A.,  Sarkar S.,    Schmidt-Hoberg K.,
  2012, JHEP, 2012, 123

\bibitem[\protect\citeauthoryear{{Gao}, {Navarro}, {Cole}, {Frenk}, {White},
  {Springel}, {Jenkins} \& {Neto}}{{Gao} et~al.}{2008}]{Gao_2008}
{Gao} L.,  {Navarro} J.~F.,  {Cole} S.,  {Frenk} C.~S.,  {White} S.~D.~M.,
  {Springel} V.,  {Jenkins} A.,    {Neto} A.~F.,  2008, MNRAS, 387, 536

\bibitem[\protect\citeauthoryear{Gao, Navarro, Frenk, Jenkins, Springel \&
  White}{Gao et~al.}{2012}]{Gao_2012}
Gao L.,  Navarro J.~F.,  Frenk C.~S.,  Jenkins A.,  Springel V.,    White S.
  D.~M.,  2012, MNRAS, 425, 2169

\bibitem[\protect\citeauthoryear{Hayashi \& White}{Hayashi \&
  White}{2008}]{Hayashi_2008}
Hayashi E.,  White S. D.~M.,  2008, MNRAS, 388, 2

\bibitem[\protect\citeauthoryear{Hernquist}{Hernquist}{1990}]{Hernquist_1990}
Hernquist L.,  1990, ApJ, 356, 359

\bibitem[\protect\citeauthoryear{Hjorth \& Williams}{Hjorth \&
  Williams}{2010}]{Hjorth_Williams_2010_I}
Hjorth J.,  Williams L.~R.,  2010, ApJ, 722, 851

\bibitem[\protect\citeauthoryear{James \& Roos}{James \& Roos}{1975}]{Minuit}
James F.,  Roos M.,  1975, Computer Physics Communications, 10, 343

\bibitem[\protect\citeauthoryear{{Jarosik}, {Bennett}, {Dunkley}, {Gold},
  {Greason}, {Halpern}, {Hill}, {Hinshaw}, {Kogut}, {Komatsu}, {Larson},
  {Limon} et~al.,}{{Jarosik} et~al.}{2011}]{WMAP_2011_A}
{Jarosik} N.,  {Bennett} C.~L.,  {Dunkley} J.,  {Gold} B.,  {Greason} M.~R.,
  {Halpern} M.,  {Hill} R.~S.,  {Hinshaw} G.,  {Kogut} A.,  {Komatsu} E.,
  {Larson} D.,  {Limon} M.,    et~al., 2011, ApJS, 192, 14

\bibitem[\protect\citeauthoryear{{Jing} \& {Suto}}{{Jing} \&
  {Suto}}{2000}]{Jing_Suto_2000}
{Jing} Y.~P.,  {Suto} Y.,  2000, ApJL, 529, L69

\bibitem[\protect\citeauthoryear{{Kang} \& {He}}{{Kang} \&
  {He}}{2011}]{Kang_He_2011_A}
{Kang} D.-B.,  {He} P.,  2011, AAP, 526, A147

\bibitem[\protect\citeauthoryear{Kang \& He}{Kang \& He}{2011}]{Kang_He_2011_B}
Kang D.-B.,  He P.,  2011, MNRAS, 416, 32

\bibitem[\protect\citeauthoryear{King}{King}{1966}]{King_1966}
King I.~R.,  1966, The Astronomical Journal, 71, 64

\bibitem[\protect\citeauthoryear{{Lapi} \& {Cavaliere}}{{Lapi} \&
  {Cavaliere}}{2011}]{Lapi_Cavaliere_2011}
{Lapi} A.,  {Cavaliere} A.,  2011, Advances in Astronomy, 2011

\bibitem[\protect\citeauthoryear{{Lemze}, {Wagner}, {Rephaeli}, {Sadeh},
  {Norman}, {Barkana}, {Broadhurst}, {Ford} \& {Postman}}{{Lemze}
  et~al.}{2012}]{Lemze_2012}
{Lemze} D.,  {Wagner} R.,  {Rephaeli} Y.,  {Sadeh} S.,  {Norman} M.~L.,
  {Barkana} R.,  {Broadhurst} T.,  {Ford} H.,    {Postman} M.,  2012, ApJ, 752,
  141

\bibitem[\protect\citeauthoryear{Lewis, Challinor \& Lasenby}{Lewis
  et~al.}{2000}]{Lewis_2000}
Lewis A.,  Challinor A.,    Lasenby A.,  2000, ApJ, 538, 473

\bibitem[\protect\citeauthoryear{Lynden-Bell}{Lynden-Bell}{1967}]{LyndenBell_1967}
Lynden-Bell D.,  1967, MNRAS, 136, 101

\bibitem[\protect\citeauthoryear{Madsen}{Madsen}{1996}]{Madsen_1996}
Madsen J.,  1996, MNRAS, 280, 1089

\bibitem[\protect\citeauthoryear{{Mamon}, {Biviano} \& {Murante}}{{Mamon}
  et~al.}{2010}]{Mamon_2010}
{Mamon} G.~A.,  {Biviano} A.,    {Murante} G.,  2010, AAP, 520, A30

\bibitem[\protect\citeauthoryear{Merritt, Graham, Moore, Diemand \&
  Terzić}{Merritt et~al.}{2006}]{Merritt_2006}
Merritt D.,  Graham A.~W.,  Moore B.,  Diemand J.,    Terzić B.,  2006, AJ,
  132, 2685

\bibitem[\protect\citeauthoryear{Merritt, Navarro, Ludlow \& Jenkins}{Merritt
  et~al.}{2005}]{Merritt_2005}
Merritt D.,  Navarro J.~F.,  Ludlow A.,    Jenkins A.,  2005, ApJ, 624,
  L85–L88

\bibitem[\protect\citeauthoryear{Mira, Hilbert, Hartlap \& Schneider}{Mira
  et~al.}{2011}]{Mira_2011}
Mira E.~P.,  Hilbert S.,  Hartlap J.,    Schneider P.,  2011, A\&A, 531, A169

\bibitem[\protect\citeauthoryear{Navarro, Frenk \& White}{Navarro
  et~al.}{1996}]{NFW_1996}
Navarro J.~F.,  Frenk C.~S.,    White S. D.~M.,  1996, ApJ, 462, 563

\bibitem[\protect\citeauthoryear{Navarro, Frenk \& White}{Navarro
  et~al.}{1997}]{NFW_1997}
Navarro J.~F.,  Frenk C.~S.,    White S. D.~M.,  1997, ApJ, 490, 493

\bibitem[\protect\citeauthoryear{Navarro, Hayashi, Power, Jenkins, Frenk,
  White, Springel, Stadel \& Quinn}{Navarro et~al.}{2004}]{Navarro_2004}
Navarro J.~F.,  Hayashi E.,  Power C.,  Jenkins A.~R.,  Frenk C.~S.,  White S.
  D.~M.,  Springel V.,  Stadel J.,    Quinn T.~R.,  2004, MNRAS, 349, 1039

\bibitem[\protect\citeauthoryear{Navarro, Ludlow, Springel, Wang, Vogelsberger,
  White, Jenkins, Frenk \& Helmi}{Navarro et~al.}{2010}]{Navarro_2010}
Navarro J.~F.,  Ludlow A.,  Springel V.,  Wang J.,  Vogelsberger M.,  White S.
  D.~M.,  Jenkins A.,  Frenk C.~S.,    Helmi A.,  2010, MNRAS, 402, 21–34

\bibitem[\protect\citeauthoryear{Rubin, Ford \& Thonnard}{Rubin
  et~al.}{1980}]{Rubin_1980}
Rubin V.~C.,  Ford W. K.~J.,    Thonnard N.,  1980, ApJ, 238, 471

\bibitem[\protect\citeauthoryear{Rubin, Thonnard \& Ford}{Rubin
  et~al.}{1978}]{Rubin_1978}
Rubin V.~C.,  Thonnard N.,    Ford W. K.~J.,  1978, ApJ, 225, L107

\bibitem[\protect\citeauthoryear{{Schmidt}, {Lima}, {Oyaizu} \& {Hu}}{{Schmidt}
  et~al.}{2009}]{Schmidt_2009}
{Schmidt} F.,  {Lima} M.,  {Oyaizu} H.,    {Hu} W.,  2009, PRD, 79, 083518

\bibitem[\protect\citeauthoryear{{S{\'e}rsic}}{{S{\'e}rsic}}{1963}]{Sersic_1963}
{S{\'e}rsic} J.~L.,  1963, Boletin de la Asociacion Argentina de Astronomia La
  Plata Argentina, 6, 41

\bibitem[\protect\citeauthoryear{Springel, White, Jenkins, Frenk, Yoshida, Gao,
  Navarro, Thacker, Croton, Helly, Peacock, Cole, Thomas, Couchman, Evrard,
  Colberg \& Pearce}{Springel et~al.}{2005}]{Springel_2005}
Springel V.,  White S. D.~M.,  Jenkins A.,  Frenk C.~S.,  Yoshida N.,  Gao L.,
  Navarro J.,  Thacker R.,  Croton D.,  Helly J.,  Peacock J.~A.,  Cole S.,
  Thomas P.,  Couchman H.,  Evrard A.,  Colberg J.,    Pearce F.,  2005,
  Nature, 435, 629

\bibitem[\protect\citeauthoryear{Stadel, Potter, Moore, Diemand, Madau, Zemp,
  Kuhlen \& Quilis}{Stadel et~al.}{2009}]{Stadel_2009}
Stadel J.,  Potter D.,  Moore B.,  Diemand J.,  Madau P.,  Zemp M.,  Kuhlen M.,
     Quilis V.,  2009, MNRAS, 398, L21

\bibitem[\protect\citeauthoryear{Sánchez, Scóccola, Ross, Percival, Manera,
  Montesano, Mazzalay, Cuesta, Eisenstein, Kazin, McBride, Mehta,
  Montero-Dorta, Padmanabhan, Prada, Rubiño-Martín \& et al.}{Sánchez
  et~al.}{2012}]{Sanchez_2012}
Sánchez A.~G.,  Scóccola C.~G.,  Ross A.~J.,  Percival W.,  Manera M.,
  Montesano F.,  Mazzalay X.,  Cuesta A.~J.,  Eisenstein D.~J.,  Kazin E.,
  McBride C.~K.,  Mehta K.,  Montero-Dorta A.~D.,  Padmanabhan N.,  Prada F.,
  Rubiño-Martín J.~A.,    et al. 2012, MNRAS, 425, 415

\bibitem[\protect\citeauthoryear{Taylor \& Navarro}{Taylor \&
  Navarro}{2001}]{Taylor_2001}
Taylor J.~E.,  Navarro J.~F.,  2001, ApJ, 563, 483

\bibitem[\protect\citeauthoryear{Tinker, Robertson, Kravtsov, Klypin, Warren,
  Yepes \& Gottlöber}{Tinker et~al.}{2010}]{Tinker_2010}
Tinker J.~L.,  Robertson B.~E.,  Kravtsov A.~V.,  Klypin A.,  Warren M.~S.,
  Yepes G.,    Gottlöber S.,  2010, ApJ, 724, 878

\bibitem[\protect\citeauthoryear{Umetsu, Broadhurst, Zitrin, Medezinski, Coe \&
  Postman}{Umetsu et~al.}{2011b}]{Umetsu_2011_b}
Umetsu K.,  Broadhurst T.,  Zitrin A.,  Medezinski E.,  Coe D.,    Postman M.,
  2011b, ApJ, 738, 41

\bibitem[\protect\citeauthoryear{Umetsu, Broadhurst, Zitrin, Medezinski \&
  Hsu}{Umetsu et~al.}{2011a}]{Umetsu_2011_a}
Umetsu K.,  Broadhurst T.,  Zitrin A.,  Medezinski E.,    Hsu L.,  2011a, ApJ,
  729, 127

\bibitem[\protect\citeauthoryear{Vikhlinin, Kravtsov, Forman, Jones,
  Markevitch, Murray \& Van~Speybroeck}{Vikhlinin
  et~al.}{2006}]{Vikhlinin_2006}
Vikhlinin A.,  Kravtsov A.,  Forman W.,  Jones C.,  Markevitch M.,  Murray
  S.~S.,    Van~Speybroeck L.,  2006, ApJ, 640, 691

\bibitem[\protect\citeauthoryear{Williams \& Hjorth}{Williams \&
  Hjorth}{2010}]{Hjorth_Williams_2010_II}
Williams L.~R.,  Hjorth J.,  2010, ApJ, 722, 856

\bibitem[\protect\citeauthoryear{Williams, Hjorth \& Wojtak}{Williams
  et~al.}{2010}]{Hjorth_Williams_2010_III}
Williams L.~R.,  Hjorth J.,    Wojtak R.,  2010, ApJ, 725, 282

\bibitem[\protect\citeauthoryear{{Zhao}}{{Zhao}}{1996}]{Zhao_1996}
{Zhao} H.,  1996, MNRAS, 278, 488

\bibitem[\protect\citeauthoryear{Zwicky}{Zwicky}{1933}]{Zwicky_1933}
Zwicky F.,  1933, Helv. Phys. Acta, 6, 110

\end{thebibliography}
\label{lastpage}
\end{document}